\documentclass[reprint,amsmath,amssymb,aps,prl,groupedaddress,nofootinbib,twocolumn,superscriptaddress]{revtex4-2}
\usepackage[english]{babel}
\makeatletter
\addto\captionsenglish{\def\l@en{\l@english}}
\makeatother

\usepackage{graphicx}
\usepackage{amssymb,amsmath,braket}
\usepackage{bm,bbm}

\usepackage{graphicx,xcolor} 
\usepackage{tabularx, booktabs} 
\usepackage{hyperref}

\usepackage{pifont}
\newcommand{\LightwaveElectronics}{THz sub-cycle lightwave}

\newcommand{\sectionPRL}[1]{%
  \refstepcounter{section}
  \phantomsection       
  \addcontentsline{toc}{section}{#1}
  \par\medskip\noindent%
  \textit{#1.---}\ %
}

\usepackage[caption=false]{subfig}
\usepackage{changes}
\definechangesauthor[name=Jonas, color=blue]{JH} 
\definechangesauthor[name=Johannes, color=red]{JK}

\begin{document}

\title{Magnetic Bloch Oscillations in Odd-Wave Magnets  \\ and the Nonlinear Edelstein Effect}

\author{Jonas Habel}
\affiliation{Technical University of Munich, TUM School of Natural Sciences, Physics Department, Garching, Germany}
\affiliation{Munich Center for Quantum Science and Technology (MCQST), Schellingstr. 4, 80799 M{\"u}nchen, Germany}
\author{Johannes Knolle}
\affiliation{Technical University of Munich, TUM School of Natural Sciences, Physics Department, Garching, Germany}
\affiliation{Munich Center for Quantum Science and Technology (MCQST), Schellingstr. 4, 80799 M{\"u}nchen, Germany}

\date{\today}

\begin{abstract}
Bloch oscillations (BOs) are a quantum phenomenon in which electrons subjected to an electric field in a  periodic potential exhibit an oscillating current without a net drift.
In real conductors, scattering reduces the coherence required for BOs driving the system toward a steady state with a DC current.
While previous studies have focused on charge transport, charge carriers also possess spin, raising the question of whether BOs can emerge in magnetic observables.
Here, we show that the magnetization of odd-wave magnets can undergo BOs before relaxing to the steady-state Edelstein value, a phenomenon we term \textit{magnetic} BOs.
Using analytical and numerical methods, we demonstrate this effect in a minimal one-dimensional model of a p-wave magnet and generalize it to two dimensions.
Our analysis further reveals that the Edelstein magnetization is generically nonlinear in the applied electric field.
Finally, we argue that magnetic BOs can be detected in materials through higher-harmonic generation in THz sub-cycle lightwave spectroscopy.
Magnetic BOs provide a genuine non-equilibrium signature of spin-charge coupling in unconventional magnets.

\end{abstract}
	
\maketitle

\sectionPRL{Introduction}
The application of a static electric field to a pristine crystalline conductor gives rise to Bloch oscillations (BOs), i.e., a collective oscillatory motion of electrons in real space without a net drift velocity.
This genuine quantum effect, discovered about a century ago by Felix Bloch~\cite{bloch1929quantenmechanik}, can be rationalized semiclassically: the quasimomentum of a wavepacket of crystal electrons, which increases linearly over time due to the acceleration by the electric field, is only defined up to a reciprocal lattice vector and therefore effectively flips its sign when exceeding the first Brillouin zone, leading to an oscillatory real-space motion.
The characteristic ``Bloch frequency" $\omega_B= eEa/\hbar$ of these oscillations is typically proportional to the electric field $E$ and the lattice constant $a$.
The fact that a static electric field causes a finite-frequency response demonstrates the inherent non-linearity of the phenomenon and is promising for technological applications~\cite{esaki1970superlattice,leo1998interband}. 

BOs were initially deemed impossible to observe in conventional bulk solids because impurities and many-body scattering with other quasiparticles generally cause the electrons to lose their coherence and relax towards a steady state with a finite drift current governed by Ohm's law.
Characteristic relaxation time scales in materials are typically shorter than the BO period for realistic static electric field strengths.
As a consequence, electrons remain close to their equilibrium distribution at all times, never traversing the edges of the Brillouin zone, thus, inhibiting BOs.
One way to overcome the restrictions is the increase of the effective lattice constant with the help of superlattice potentials, which enabled the first observation of BO in semiconductor heterostructures~\cite{feldmann1992optical,leo1998interband}.
This result has been of fundamental importance as it settled the long-standing question of the observability of BOs~\cite{leo1998interband}.
Furthermore, it enables the generation of THz electromagnetic field radiation from the application of DC electric fields~\cite{waschke1993coherent}.
Recently, the flat band regime of twisted Moire systems has been proposed as an ideal setting for observing BOs~\cite{fahimniya2021synchronizing,de2023berry,de2024floquet}.    

Another way to go beyond the scattering restrictions is to increase the electric field intensity using time dependent fields. The advent of Terahertz spectroscopy has enabled the fabrication of high-intensity laser pulses with significantly stronger electric field amplitudes than previously possible, which -- albeit being AC fields -- are able to drive the electrons far from equilibrium before they scatter.
The resulting BO-like electron motion manifests in the response spectrum of the material as higher harmonics of the driving laser frequency, a phenomenon known as \textit{higher-harmonic generation} (HHG)~\cite{ghimire2011observation,schubert_sub-cycle_2014,hohenleutner2015real}.
Importantly, Bloch oscillations are a single-band phenomenon.
In multiband systems, they coexist with interband tunneling, giving rise to complex HHG spectra.
Beyond solid state systems, BOs have been demonstrated in other settings like optical waveguides~\cite{peschel1998optical} and macroscopic mechanical simulators~\cite{neder2024bloch}. Most prominently, they have been observed in optical lattices of cold atomic gases~\cite{dahan1996bloch} where they can be used to probe interaction effects~\cite{gustavsson2008control} and various forms of band topology~\cite{price2012mapping,dauphin2013extracting,atala2013direct,aidelsburger2015measuring}.

\begin{figure*}
    \centering
    \includegraphics[width=0.9\linewidth]{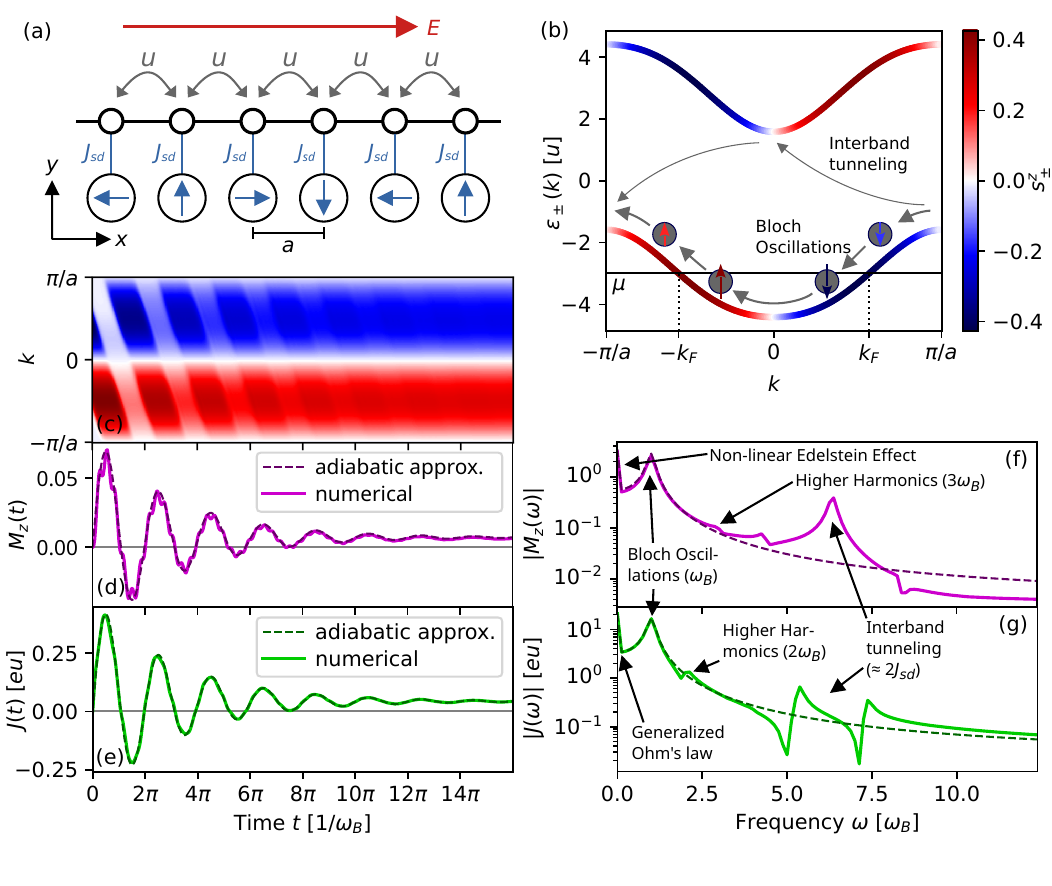}
    \caption{
    (a) Sketch of the minimal 1D model proposed in~\cite{brekke_minimal_2024}. Electrons can hop to neighboring lattice sites with spin-independent hopping amplitude $u$. The electron spins are coupled to local magnetic moments via a Kondo-type interaction $J_{sd}$. The local moments form a $90^\circ$ spiral in the $xy$-plane that breaks inversion symmetry. At time $t=0$, an electric field $E$ is switched on along the $x$-direction.
    (b) Electronic band dispersion in the first lattice Brillouin zone with chemical potential $\mu = -3u$ and $J_{sd} = 3u$. The color indicates the $z$-projection of the electron spin. An electron wavepacket in the lower band accelerated by the electric field undergoes linear motion in momentum space giving rise to in-band (magnetic) Bloch oscillations (thick arrows) and interband tunneling (thin arrows).
    (c) Spin-weighted electron distribution function of the lower band vs.\ time and momentum from numerical solution of the Boltzmann equation~\eqref{eq:Boltzmann}.
    (d) Total magnetization of the system vs.\ time from numerical solution (solid line) and analytic adiabatic approximation (dashed line) of the Boltzmann equation~\eqref{eq:Boltzmann}.
    (e) Numerical and analytic results for the charge current vs.\ time.
    Panels (c-e) use $\mu = -3u$, $J_{sd} = 3u$, $\omega_B = u$ and $\tau = 10/u$. The initial Bloch oscillations are disrupted by scattering, leading to a steady-state distribution with finite magnetization (Edelstein effect).
    (f) Fourier spectrum of the magnetization from panel (c).
    (g) Fourier spectrum of the current from panel (d).
    Panels (f-g) show both the numerical (solid lines) and analytic results (dashed lines).
    }
    \label{fig:fig1}
\end{figure*}

The discussion of BOs has focused on the out-of-equilibrium response of the \textit{charge} degrees of freedom (d.o.f.) to an electric field, neglecting the spin d.o.f. and the dynamics of observables like the magnetization. 
This assumption is reasonable in conventional conductors because only small relativistic effects of spin orbit coupling intertwine charge and spin.
However, recent years have seen increased interest in itinerant antiferromagnets with spin split bands like altermagnets~\cite{vsmejkal2022emerging,jungwirth2025altermagnetism,jungwirth2026symmetry}. Of particular importance are non-centrosymmetric antiferromagnets like certain non-collinear systems~\cite{hayami2020spontaneous,naka2019spin,gonzalez2024non} and the recently proposed odd-wave magnets~\cite{chakraborty2025highly,leeb2026collinear}, which can exhibit a strong Edelstein effect even in the absence of SO coupling~\cite{aronov1989nuclear,edelstein1990spin,sinova2015spin}. There, an applied electric field leads to a steady-state magnetization perpendicular to the field direction despite the system having zero net magnetization in equilibrium.
This Edelstein effect can be sizable due to large non-relativistic spin splitting in odd-wave magnets, which raises the natural question of how the magnetization evolves in the short-time regime after application of an electric field and how it relaxes to the steady state Edelstein value.

In this paper, we demonstrate that the {\it magnetization} of an itinerant $p$-wave magnet can undergo Bloch oscillations, analogous to the charge current.
The necessary conditions for the occurrence of such {\it magnetic BOs} are: (i) the breaking of inversion symmetry (inherent to any odd-wave magnet); (ii) the breaking of spin conservation so that the system can develop imbalanced spin populations when driven out of equilibrium; and (iii) the usual BO requirement of the period  $2\pi/\omega_B$ being smaller than the relaxation time.
We calculate the full time-dependence of  magnetic BOs and recover the steady-state Edelstein magnetization in the long time limit.
Importantly, this approach also captures the nonlinear response that emerges in the strong-field regime.
Generalizing the setting to time dependent AC electric fields, we propose to measure magnetic BOs with \LightwaveElectronics{} spectroscopy.
The resulting HHG of THz oscillations of the magnetization can be measured via the time-resolved Kerr or Faraday rotation~\cite{kampfrath2011coherent,kimel2009inertia,hudl2019nonlinear}.

\begin{figure*}[h]
    \centering
    \includegraphics[width=0.9\linewidth]{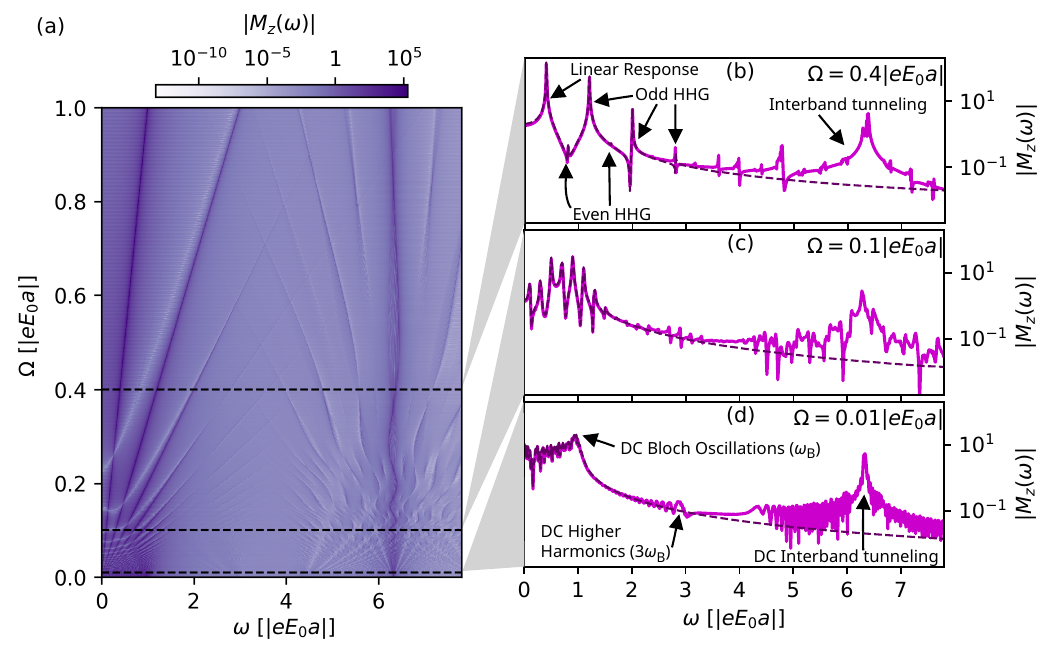}
    \caption{(a) Fourier transform of the magnetization vs.\ frequency $\omega$ and the external driving frequency $\Omega$ from a numerical solution of the Boltzmann equation~\eqref{eq:Boltzmann} for $u = |eE_0a|$, $J_{sd} = 3u$, $\mu = -3u$, and $\tau = 1000/u$.
    Dashed lines indicate the constant-$\Omega$ cuts depicted in panels (b-d).
    The latter show both the numerical solution (solid lines) and analytical adiabatic approximation (dashed lines).
    In panel (d), $\omega_B = eE_0a$ denotes the Bloch frequency in the DC limit $\Omega \to 0$.
    }
    \label{fig:fig2}
\end{figure*}

\sectionPRL{Minimal model} We first illustrate the main idea of magnetic BOs within a minimal model of electrons in 1D coupled to a helical texture of local magnetic moments, see Fig.~\ref{fig:fig1}(a), governed by the double-exchange tight-binding Hamiltonian~\cite{brekke_minimal_2024}
\begin{align}
\label{eq:H_1D}
    \hat H = &-u \sum_{\langle ij \rangle,\sigma} ( c_{i\sigma}^\dagger c_{j\sigma} + h.c.) \nonumber \\
    & -~J_{sd} \sum_{j,\sigma\sigma'} \vec S_j \cdot c_{j\sigma}^\dagger \vec \sigma_{\sigma\sigma'} c_{j\sigma'},
\end{align}
where $c_{j\sigma}^\dagger$ creates an electron at site $j$ with spin $\sigma$, $u$ denotes the hopping strength between nearest neighbor sites $\langle ij \rangle$, $J_{sd}$ is the sd-type coupling between the itinerant electrons and the local moments, which are given by $\vec S_j = (\cos(\pi j/2), -\sin(\pi j/2), 0)$.
Note that we fix these moments to serve as a classical background field.
Also, $\vec\sigma = (\sigma_x, \sigma_y, \sigma_z)$ is the vector of Pauli matrices, $a$ is the lattice constant, and we use the convention $\hbar \equiv 1$.
The Hamiltonian~\eqref{eq:H_1D} breaks $[\mathcal T || \mathcal T \mathcal P \vec t]$ and $[C_2^{x/y} || \vec t]$ symmetry for any lattice translation $\vec t$, allowing for spin-split electronic bands~\cite{leeb2026collinear} (here, $\mathcal T$, $\mathcal P$ and $C_2^{x/y}$ denote time reversal, real-space inversion and a $\pi$-rotation in spin space about the $x/y$-axis, respectively).
However, it preserves $[\mathcal T || \mathcal T \vec t_2]$ symmetry (with $\vec t_2$ a real-space translation by two sites), which protects the $p$-wave spin polarization of the bands.
Also note that the Hamiltonian breaks spin conservation so that the bands cannot be labeled by a spin quantum number.

The system possesses a hidden $[ C_4^z || \vec t_1 ]$ symmetry (where $C_4^z$ and $\vec t_1$ denote a spin-space rotation by $\pi/2$ about the $z$-axis and a real-space translation by one site, respectively), which allows us to reduce the four site unit cell down to one, resulting in two bands. To do so, we perform a site-dependent rotation of the electronic spin such that $\vec S_j$ is aligned with the local $z$-axis.
The transformed fermionic operators are $d_{j\tau} = \sum_{\sigma} \left(U_j\right)_{\tau\sigma} c_{j\sigma}$ with $U_j$ the corresponding unitary matrix at site $j$, see Supplementary Material (SM).
In the rotated basis, the Hamiltonian has a manifestly single-site unit cell.
Using the Fourier transform $d_{j\tau} = N^{-1/2} \sum_k e^{ikja} d_{k\tau}$, the Hamiltonian takes the form $\hat H = \sum_{k,\tau\tau'} d_{k\tau}^\dagger h_{\tau\tau'}(k) d_{k\tau'}$, where $h(k)$ is the $2\times 2$ Bloch Hamiltonian.
Diagonalizing $h(k) = \mathcal U(k) D(k)\, \mathcal U^\dagger(k)$ with $D(k) = \operatorname{diag}(\epsilon_+(k), \epsilon_-(k))$ results in two bands that reside in a fourfold enlarged Brillouin zone, with dispersion $\epsilon_\pm(k) = -\sqrt 2\, u\cos(ka) \pm \sqrt{J_{sd}^2 + 2u^2 \sin^2(ka)}$ and $p$-wave spin polarization $s_{\pm}^z(k) = \pm \sqrt 2\,u\sin(ka) \big/ \sqrt{J_{sd}^2 + 2u^2 \sin^2(ka)}$ as illustrated in  Fig.~\ref{fig:fig1}(b).
The spin polarizations in $x$ and $y$ direction vanish after averaging over the original four-site unit cell. Notably, 
the strength of the sd-coupling $J_{sd}$ provides a tuning knob for the band gap.

\sectionPRL{Boltzmann transport theory}
\label{sec:Boltzmann_transport_theory}
In equilibrium, the model exhibits zero net magnetization because every filled single-particle energy level has a time-reversal-related partner at the same energy with the opposite spin polarization.
However, an applied electric field $E$ can break this balance and lead to a finite net magnetization $M_z$ in the steady state due to the lack of an inversion center. Our objective is to study the full magnetization dynamics including relaxation to the steady state. We use transport theory treating the position and momentum degrees of freedom semiclassically but retain the quantum dynamics in the spin/band degree of freedom.
This allows us to treat and differentiate both intraband Bloch oscillations and interband tunneling.

The main quantity of interest is the electron distribution function $f(k, t)$, which is a hermitian $2\times 2$-matrix in the local rotated spin basis (i.e., the basis of the $d_{k\tau}$ fermions).
Here, we assume a homogeneous distribution such that $f$ does not depend on the spatial coordinate.
Its time evolution is governed by the spinor Boltzmann equation
\begin{align}
\label{eq:Boltzmann}
    \frac{\partial f}{\partial t} - eE \frac{\partial f}{\partial k} + i [h(k), f] = I[f],
\end{align}
where $I[f]$ denotes the collision integral.
Note that, without the collision integral, this equation is in fact equivalent to the full unitary quantum time evolution of the one-particle density matrix, see SM.
To enable analytical progress, we adopt the relaxation time approximation $I[f] = -(f - f^\mathrm{eq}) / \tau$, which collects all microscopic spin-isotropic scattering and relaxation processes towards the equilibrium distribution $f^\mathrm{eq}$ into a single phenomenological relaxation time scale $\tau$.
This standard approximation provides an efficient interpolation between the short-time out-of-equilibrium dynamics and the long-time steady state. We neglect spin-flip scattering. 

To systematically identify and separate the energy scales involved in the intraband Bloch oscillations and the interband tunneling, we transform the Boltzmann equation~\eqref{eq:Boltzmann} to band space:
\begin{align}
\label{eq:Boltzmann_bandspace}
    \frac{\partial\tilde f}{\partial t} - eE \frac{\partial\tilde f}{\partial k} + i [D(k), \tilde f] + i eE [\mathcal A(k), \tilde f] = \tilde I[\tilde f]
\end{align}
where $\tilde f(k, t) \equiv \mathcal U^\dagger(k) f(k, t)\, \mathcal U(k)$ and $\tilde I[\tilde f] = -(\tilde f - \tilde f^\mathrm{eq}) / \tau$ are the distribution function and the collision integral in band space, respectively, and $\mathcal A(k) \equiv i\mathcal U^\dagger(k) \partial\mathcal U(k) / \partial k$ is the non-abelian Berry connection.
Since $D(k)$ is diagonal, the only term in \eqref{eq:Boltzmann_bandspace} that can lead to interband transitions is the one involving $\mathcal A(k)$.
Notably, $\mathcal A(k)$ is inversely proportional to the direct band gap, which is set by the sd-coupling $J_{sd}$.
Therefore, the interband transition term scales as $\omega_B/J_{sd}$, where $\omega_B = eEa$ is the Bloch frequency.
In the limit of well-separated bands, $u \ll J_{sd}$ and comparatively weak fields, $\omega_B \ll J_{sd}$, the time evolution is confined to the individual bands, which we refer to as the \textit{adiabatic} approximation.

First, we solve the Boltzmann equation analytically in the adiabatic approximation without relaxation ($\tau \to \infty$) for the scenario that the system is prepared in a zero-temperature equilibrium state with the external electric field switched on at time $t=0$.
For simplicity, we assume that the bands are well separated, $J_{sd} \gtrsim u, \omega_B$, and that the chemical potential $\mu$ lies inside the lower band, $\mu = -J_{sd}$.
The initial equilibrium distribution is then given by $\tilde f^\mathrm{eq}(k) = \operatorname{diag} (0, \Theta(\mu - \epsilon_-(k)))$, with the Heaviside step function~$\Theta$.
The time-evolved distribution function takes the form $\tilde f(k, t) = \tilde f^\mathrm{eq}(k + eEt)$, corresponding to a linear growth of the crystal momentum with time.
The charge current in response to the applied field is $J(t) = \sqrt 2\,\pi e u \sin(k_F a) \sin(\omega_B t)$ and performs perpetual Bloch oscillations with frequency $\omega_{B} = eEa$, as expected.  
By transforming this solution back into the lab frame of the operators $c_{i\sigma}$ and averaging over the four-site unit cell, we obtain the time-dependent net magnetization $M_z(t) = (\sqrt 2\, \pi)^{-1} (u / J_{sd}) \sin(k_F a) \sin(\omega_B t)$, where $k_F\ge 0$ is the Fermi momentum. Remarkably, also the magnetization undergoes BO with the very same frequency! 
Note that the Bloch oscillations are insensitive to the physical four-site unit cell since the dominant BO frequency $eE \cdot a$ is that of the single-site unit cell with lattice constant $a$.
This is a result of the hidden $[C_4^z || \vec t_1]$ symmetry, see SM for discussion.

Second, we include a finite relaxation time $\tau$ in the adiabatic approximation to describe the inevitable damping of BOs.
The analytic expression for the magnetization and the charge current in this case are given in the SM  \eqref{eq:SM_magn_adiabatic} and \eqref{eq:SM_curr_adiabatic}, respectively.
Fig.~\ref{fig:fig1} presents the analytic solution for the magnetization, panel (d), and the charge current, panel (e), versus time (dashed purple and green lines, respectively) alongside the results obtained from a numerical solution of \eqref{eq:Boltzmann} (solid purple and green lines). 
Panels (f) and (g) show the corresponding Fourier spectra.
The damped Bloch oscillations with frequency $\omega_B$ manifest as broadened peaks at that frequency in Fourier space.
Furthermore, we observe weak higher-harmonic peaks in the numerical solutions at $3\omega_B$ in the magnetization and $2\omega_B$ in the current, which are caused by deviations of the band dispersion from a pure cosine.
They appear only at higher order in $u/J_{sd}$ and are thus not reproduced by our analytic treatment in the adiabatic approximation.
Note that, in contrast to the current, the magnetization contains only odd higher harmonics.
Although this may suggest a possible experimental distinction between the two responses, we caution that the higher-harmonic spectrum is not universal and depends sensitively on the details of the model (see SM).

Additionally, the numerical Fourier spectra exhibit signatures of interband tunneling at frequencies near the direct band gap, $\approx 2J_{sd}$.
Due to their non-adiabatic nature, they are of first or higher order in $\omega_B/J_{sd}$ and not captured by the analytical calculation in the adiabatic approximation.
At first order in $\omega/J_{sd}$, the dominant interband tunneling peak frequency can be computed analytically to all orders in $u/J_{sd}$, yielding a value of $(4J_{sd}/\pi) E(-2u^2/J_{sd}^2)$ with $E$ the complete elliptic integral of the second kind, see SM.
Finally, at times much larger than $\tau$, the magnetization approaches its Edelstein value of $M_z(t \gg \tau) = (\sqrt 2\,\pi)^{-1} (u / J_{sd}) \sin(k_F a)\, \omega_B\tau / (1 + (\omega_B\tau)^2)$ in the steady state, as evidenced by the zero-frequency peak in the Fourier-transform of the magnetization.
Notably, the Edelstein magnetization is nonlinear in the electric field.
For strong fields compared to the scattering rate, $\omega_B\tau \gtrsim 1$, the steady state magnetization decreases with higher field, manifesting in a negative differential Edelstein susceptibility $\chi(E) = \partial M_z/\partial E$.
The physical reason for this is the stronger Wannier-Stark localization of the electrons as the field increases.
Note that this argument only holds in the adiabatic approximation where $\omega_B \ll J_{sd}$.
When the field becomes of the order of the direct band gap, interband transitions become possible, allowing the electrons to tunnel out of their Wannier-Stark localization region and yielding further linear and nonlinear contributions to the Edelstein magnetization $M_z(t \gg \tau)$.

\sectionPRL{Experimental Probes and Higher-Harmonic Generation}
The main challenge in measuring Bloch oscillations experimentally is to drive the band electrons far out of equilibrium by accelerating them fast enough to cross the Brillouin zone boundary before they undergo scattering~\cite{feldmann1992optical,peschel1998optical}.
With the advent of ultrafast \LightwaveElectronics{} spectroscopy, it has become possible to generate electric fields that are strong enough to provide this acceleration~\cite{reislohner_onset_2022,reimann2018subcycle,schmid_tunable_2021}.
Since these driving fields are produced by Terahertz lasers, our previous assumption of a static electric field in the theoretical description is no longer valid and must be replaced with an oscillatory field $E(t) = E_0 \cos(\Omega t)$, where $E_0$ is the field amplitude and $\Omega$ is the driving frequency.
In the adiabatic approximation without relaxation, the Boltzmann equation can again be solved exactly, giving $M_z(t) = (\sqrt 2\,\pi)^{-1} (u / J_{sd}) \sin(k_F a) \sin(eE_0a/\Omega \sin(\Omega t))$.
At weak driving fields $eE_0a \ll \Omega$, the magnetization oscillates at the driving frequency, $M_z(t) \propto \sin(\Omega t)$, as predicted by linear response theory.
At stronger driving fields, higher orders in $eE_0a/\Omega$ become important, which contain oscillatory terms of the form $\sin(n\Omega t)$ with odd-integer $n$.
These show up as odd harmonics of the driving frequency in the response spectrum (HHG), which are a tell-tale sign of the electron distribution exploring a large region of the Brillouin zone analogous to the magnetic Bloch oscillations in the DC field case (see Fig.~\ref{fig:fig2}).
Their amplitudes are suppressed by a factor of $J_n(eE_0 a/\Omega)$ (where $J_n$ is the $n$-th Bessel function of the first kind), which reduces to the known perturbative scaling law $\propto E_0^n$ at intermediate field strengths $eE_0a \lesssim \Omega$~\cite{schmid_tunable_2021}.
Note that the phenomenon of HHG only refers to higher harmonics in the \emph{driving} frequency $\Omega$ of the AC field, which is distinct from the higher harmonics in the \emph{Bloch} frequency $\omega_B$ in the presence of a DC field as discussed in the previous section.
The inclusion of a finite relaxation time leads to the emergence of even harmonics of the driving frequency, while interband tunneling gives rise to further spectral signatures at frequencies comparable with the band gap.
In the DC limit $\Omega \ll |eE_0a|$, the harmonics become very dense in the spectrum, bunching up near the DC Bloch freqeuency $\omega_B = eE_0 a$, its higher harmonics, and the DC interband tunneling frequency (see Fig.~\ref{fig:fig2}(d)), eventually reproducing Fig.~\ref{fig:fig1}(f) as $\Omega \to 0$.

Observing {\it magnetic} BOs experimentally requires a time-resolved measurement of the magnetization.
We propose to measure the HHG spectrum exploiting the Faraday effect and/or the magnetoelectric Kerr effect on the THz probe pulse~\cite{kampfrath2011coherent,kimel2009inertia,hudl2019nonlinear}.
This will enable a clear experimental distinction between the contributions from magnetic BOs and the charge current BOs.

\begin{figure*}[h]
    \centering
    \includegraphics[width=0.9\linewidth]{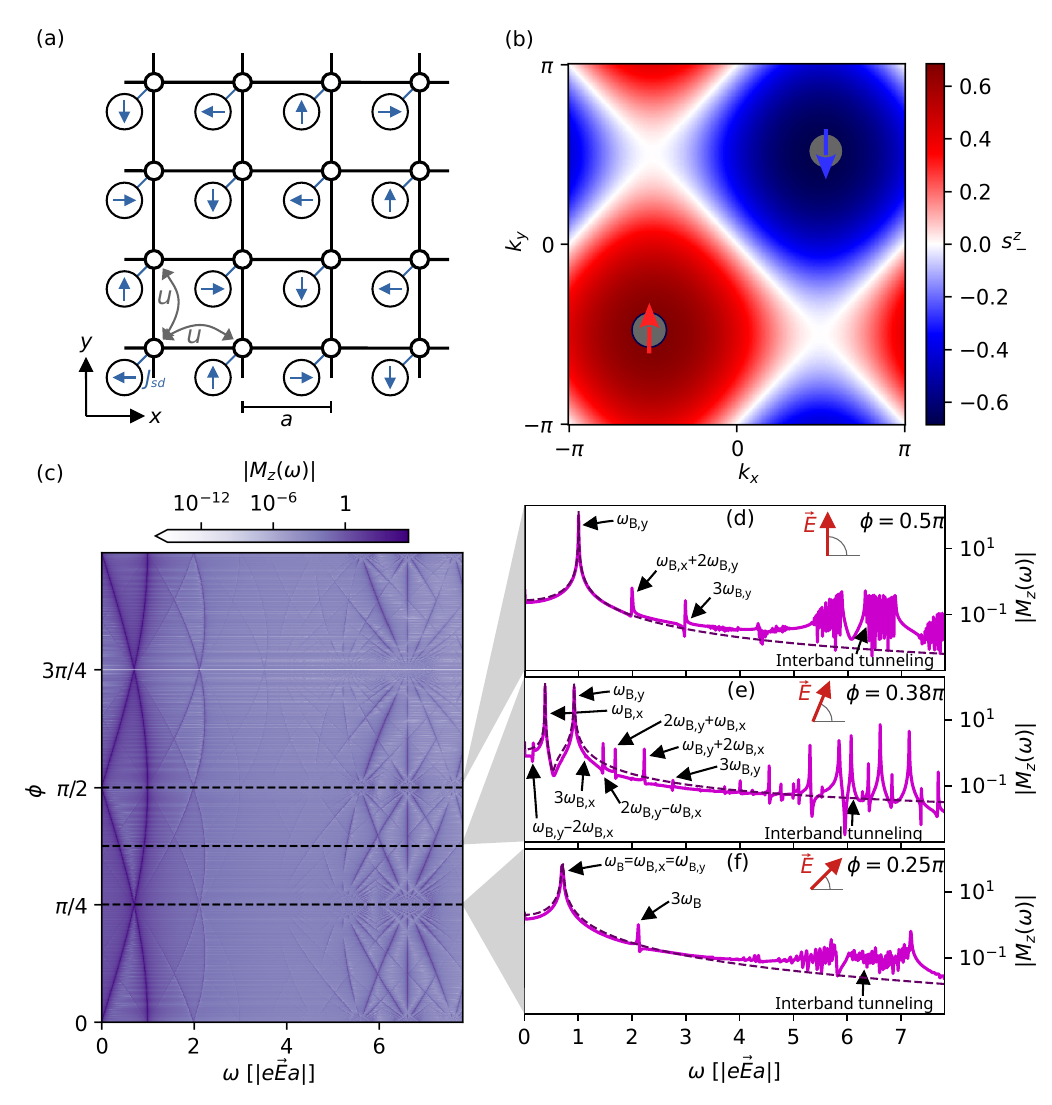}
    \caption{
    (a) Sketch of the 2D square lattice generalization of the minimal 1D model, analogous to Ref.~\cite{brekke_minimal_2024}. Electrons can hop to neighboring lattice sites with spin-independent hopping amplitude $u$. The electron spins are coupled to local magnetic moments via a Kondo-type interaction $J_{sd}$. The local moments form a $90^\circ$ spiral along both the $x$- and $y$-directions that breaks inversion symmetry.
    (b) $z$-projection of the electron spin in the lower band across the first lattice Brillouin zone for $J_{sd} = 3u$.
    (c) Fourier transform of the magnetization vs.\ frequency $\omega$ and angle $\phi$ between the electric field and the $x$-axis from a numerical solution of the Boltzmann equation~\eqref{eq:Boltzmann} for $u = \omega_B$, $J_{sd} = 3u$, $\mu = -3u$, and $\tau = 1000/u$.
    The magnetization vanishes at $\phi = 3\pi/4$ due to the mirror symmetry along the $(1,1)$ direction at that field angle.
    Dashed lines indicate the constant-$\phi$ cuts depicted in panels (d-f).
    The latter show both the numerical solution (solid lines) and analytical adiabatic approximation (dashed lines).
    Depending on the field angle, there are distinct signatures of both Bloch frequencies $\omega_{B,x}$ and $\omega_{B,y}$, as well as their higher harmonics.
    Due to the large relaxation time $\tau$, the Edelstein effect, represented by the $\omega=0$ component of the Fourier spectrum, is significantly smaller than in Fig.~\ref{fig:fig1}.
    }
    \label{fig:fig3}
\end{figure*}

\sectionPRL{Two-dimensional generalization}
To demonstrate the generality of the magnetic Bloch oscillations, we extend the above minimal model to two dimensions.
There are many ways to do this, but all of them lead to magnetic Bloch oscillations as long as $[\mathcal T || \mathcal T \mathcal P \vec t]$ is preserved and spin conservation is broken.
For illustration, we focus on a square lattice where each row is a copy of the above 1D model and the local moments between adjacent rows are twisted by $90^\circ$ with respect to each other, see Fig.~\ref{fig:fig3}(a).
The resulting 2D model is equivalent to a special case of the model proposed in Ref.~\cite{brekke_minimal_2024}.
The hidden symmetries $[C_4^z || \vec t_1^x]$ and $[C_4^z || \vec t_1^y]$ (with $\vec t_1^{x/y}$ denoting a translation by one site in the $x/y$ direction) again enable the reduction to an effective two-band system.
The spin polarization of the lower band is $s^z_-(k_x, k_y) = -\sqrt 2\,u S(k_x, k_y) / \sqrt{J_{sd}^2 + 2u^2 S(k_x, k_y)^2}$, where $S(k_x, k_y) = \sin(k_x a) + \sin(k_y a)$ (see Fig.~\ref{fig:fig3}(b)).
In two dimensions, the direction of the electric field $\vec E$ is an additional tuning parameter, which we characterize by the in-plane angle $\phi$ between the field and the $x$-axis.
In Fig.~\ref{fig:fig3}(c-f), we plot the Fourier spectra of the magnetization at various angles $\phi$ for the representative parameter choice $u = e|\vec E|a$ and $J_{sd} = 3u$.
In contrast to the 1D model, there are now two distinct Bloch frequencies $\omega_{B,x} = e|\vec E|a \cos(\phi)$ and $\omega_{B,x} = e|\vec E| a \sin(\phi)$. 
Additionally, the higher harmonic contributions in $u/J_{sd}$ due to deviations of the band dispersion from pure cosines can mix the two Bloch frequencies, yielding peaks not only at multiples of the individual Bloch frequencies but also at their sum and difference frequencies.
As in the 1D model, the higher-harmonic spectrum is non-universal.
In our case, we happen to obtain only odd higher harmonics, that is, at frequencies $n_x \omega_{B,x} + n_y \omega_{B,y}$ with integers $n_x, n_y$ whose sum is odd.
The spectral features at around twice the band gap ($\approx 6e|\vec E|a$) are again due to interband transitions.
A noteworthy feature is that the magnetization signal vanishes completely at $\phi = 3\pi/4$ because of symmetry reasons.
If the field is applied at that angle, the system possesses the symmetry $[C_2 || \sigma_{(1,1)}]$ (where $\sigma_{(1,1)}$ is a real-space reflection about the $(1, 1)$-axis), which enforces the vanishing of the magnetic Bloch oscillations and the Edelstein magnetization, which could serve as a clear experimental signature.

\sectionPRL{Conclusion and Outlook} 
We have shown that Bloch oscillations in odd-wave magnets can arise not only in the charge current but also in the magnetization due to the inversion-asymmetric spin polarization of the bands. These oscillations can be directly detected via HHG in \LightwaveElectronics{} spectroscopy.
Moreover, we have demonstrated that the Edelstein magnetization in the steady state generally contains contributions which are nonlinear in the applied electric field.
Our results highlight a non-equilibrium signatures of unconventional magnets which might be of potential technological relevance as a DC electric field is converted to an AC magnetic oscillation.
The symmetry arguments and model calculations we provide can be extrapolated to $f$-wave and higher odd-wave magnets, as well as to spin-orbit coupled systems.
For the latter, however, both BOs and the Edelstein effect are generally expected to be weak due to the relative smallness of the relativistic spin splitting.

Our work points to a range of future research questions.
We expect BOs to manifest in a wide range of observables extending beyond the magnetization.
For instance, while even-wave altermagnets do not exhibit the linear Edelstein effect due to inversion symmetry, they do host polarized spin currents in the steady state under an applied electric field.
We expect these spin currents to similarly undergo BOs in the out-of-equilibrium regime.
In the present work, we have treated the helical magnetic texture as a static background field but it is expected to have its own dynamics.
Understanding the coupled nonlinear charge and spin dynamics will be a formidable challenge.
Due to the direct coupling between the local magnetization to the charge current, we expect novel `active matter' type dynamics of domain walls~\cite{hardt2025propelling}.


\begin{acknowledgments}
\sectionPRL{Acknowledgments} We thank Christian Back and Achim Rosch for stimulating discussions. 
J.H. thanks Peng Rao for helpful discussions.
J.K. acknowledges support from the Deutsche Forschungsgemeinschaft (DFG, German Research Foundation) under grants TRR 360 - 492547816, KN1254/1-2, KN1254/2-1 and under Germany’s Excellence Strategy EXC-2111-390814868, as well as the Munich Quantum Valley, which is supported by the Bavarian state government with funds from the High- tech Agenda Bayern Plus. J.K. further acknowledges support from the Imperial-TUM flagship partnership and the Keck foundation.
\end{acknowledgments}

%

\clearpage

%

\onecolumngrid
\begin{center}
\textbf{\large Supplemental Material}
\end{center}
\twocolumngrid
\appendix

\sectionPRL{\thesection.~Minimal 1D model}
\label{sec:SM_minimal_1D_model}
To diagonalize the Hamiltonian in \eqref{eq:H_1D}, we exploit the combined $\pi/2$ spin rotation and single-site translation symmetry $[C_4^z || \vec t_1]$ by rotating the electron spin quantization axis on each site $j$ so that it aligns with the local magnetic moment.
We define the fermion operators in this rotated spin basis as $d_{j\tau} = (U_j)_{\tau\sigma} c_{j\sigma}$, where $U_j$ is the $2\times 2$ matrix corresponding to the spin-$1/2$ representation of the spin rotation (repeated spin indices are implicitly summed over).
It obeys $(U_j)_{\tau\sigma} \vec\sigma_{\sigma\sigma'} (U_j^*)_{\sigma'\tau'} = (R_j \vec\sigma)_{\tau\tau'}$ ($\dagger$), where
\begin{align}
\label{eq:SM_spin_rot_mat}
    R_j = \begin{pmatrix}
        \sin(\pi j/2) & 0 & \cos(\pi j/2) \\
        \cos(\pi j/2) & 0 & -\sin(\pi j/2) \\
        0 & 1 & 0
    \end{pmatrix}
\end{align}
is the vector representation of the spin rotation.
By design, $R_j$ rotates the local moment on site $j$ towards the $z$-axis, that is, $\vec S_j^T R_j = (0, 0, 1)$ ($\dagger\dagger$).
Expressing the local s-d coupling in \eqref{eq:H_1D} in terms of the $d$-fermions and using the relations $(\dagger)$ and $(\dagger\dagger)$, we get $-J_{sd} \sum_j d_{j\tau}^\dagger (\sigma_z)_{\tau\tau'} d_{j\tau'}$.
The hopping term in \eqref{eq:H_1D} becomes $-u \sum_{\langle ij \rangle} d_{i\tau}^\dagger (U_i)_{\tau\sigma} (U_j^*)_{\sigma\tau'} d_{j\tau'}$.
Finally, we evaluate the product $U_i U_j^\dagger = (\mathbbm 1 - i\sigma_y)/\sqrt{2}$ so that, in total,
\begin{align}
    \hat H = &-\frac{u}{\sqrt 2} \sum_{\langle ij \rangle} \left( d_i^\dagger (\mathbbm 1 - i\sigma_y) d_j + h.c. \right) \nonumber \\
    & -~J_{sd} \sum_j d_j^\dagger \sigma_z d_j
\end{align}
with $d_j \equiv (d_{j\uparrow}, d_{j\downarrow})^T$.
This Hamiltonian has a manifest single-site unit cell instead of the four-site unit cell of \eqref{eq:H_1D}.
By performing the Fourier transformation $d_j = N^{-1/2} \sum_j e^{ikja} d_k$ with momentum $-\pi/a < k \le \pi/a$ in the single-site Brillouin zone, we find the momentum space Hamiltonian
\begin{align}
    \hat H &= \sum_k d_k^\dagger h(k) d_k, \text{ where} \\
    h(k) &= -\sqrt 2\, u \cos(ka) - \sqrt 2\, u \sin(ka)\, \sigma_y - J_{sd} \sigma_z.
\end{align}
The Bloch Hamiltonian $h(k)$ is diagonal in the rotated fermion basis $\alpha_k \equiv (\alpha_{k,+}, \alpha_{k,-})^T = \mathcal U^\dagger(k) d_k$, where
\begin{align}
    \mathcal U(k) = \begin{pmatrix}
        \cos\frac{\vartheta_k}{2} & -\sin\frac{\vartheta_k}{2} \\
        i \sin\frac{\vartheta_k}{2} & i \cos\frac{\vartheta_k}{2}
    \end{pmatrix}
\end{align}
with $\tan\vartheta_k = \sqrt 2\,u \sin(k a) / J_{sd}$, yielding two energy bands with the dispersion $\epsilon_\pm(k)$ given in the main text.
Note that these two bands, when folded back into the four-site Brillouin zone, yield the eight bands expected in the original model \eqref{eq:H_1D}.
The spin polarization $\vec s_\pm$ is obtained by averaging the spin expectation value of the Bloch states over the four-site unit cell:
\begin{align}
    &\vec s_\pm(k) = \frac{1}{4} \sum_{j=1}^4 \braket{0 | \alpha_{k,\pm} (R_j \vec\sigma) \alpha_{k,\pm}^\dagger | 0} \\
    &= \pm \frac{1}{4} \sum_{j=1}^4 \begin{pmatrix} 0 \\ 0 \\ \sin\vartheta_k \end{pmatrix}
    = \frac{\pm \sqrt 2\,u \sin(k a)}{\sqrt{2u^2 \sin^2(k a) + J_{sd}^2}} \begin{pmatrix}
        0 \\ 0 \\ 1
    \end{pmatrix}, \nonumber
\end{align}
where $\ket 0$ is the vacuum state and we used that $\braket{0 | \alpha_{k\pm} \vec\sigma \alpha_{k\pm}^\dagger | 0} = (0, \sin\vartheta_k, \cos\vartheta_k)^T$.
These results agree with~\cite{brekke_minimal_2024} (up to a global sign because we chose a clockwise spiral for the local moments instead of an anticlockwise one like in~\cite{brekke_minimal_2024}).
However, the benefit of our derivation is that it only requires the diagonalization of a $2 \times 2$ instead of an $8\times 8$ Bloch Hamiltonian.

\sectionPRL{\thesection.~Boltzmann equation}
\label{sec:SM_Boltzmann_eq}
We derive the spinor Boltzmann equation in the homogeneous case without collision integral (\eqref{eq:Boltzmann}) directly from the quantum time evolution of the one-particle density matrix.
The full Hamiltonian of the minimal model with electric field is
\begin{align}
    \hat H_\mathrm{full} &= \hat H + eEa \sum_j j d_j^\dagger d_j \\
    &= \hat H + ieE \sum_k d_k^\dagger \frac{\partial}{\partial k} d_k.
\end{align}
(Spin indices are suppressed for readability.)
We define the one-particle density matrix in the rotated spin basis as $\rho_{\tau\tau'}(k, t) = \braket{d_{k\tau'}^\dagger(t) d_{k\tau}(t)}$, where $\hat O(t) = e^{-i\hat H_\mathrm{full} t} \hat O e^{i\hat H_\mathrm{full} t}$ denotes the Heisenberg time evolution of an operator $\hat O$ under the full Hamiltonian.
Next, we use the Heisenberg equation of motion to find that
\begin{align}
    &\frac{\partial \rho_{\tau\tau'}(k, t)}{\partial t} = i \braket{\left[ \hat H_\mathrm{full}, d_{k\tau'}^\dagger d_{k\tau} \right]\!(t)} \\
    &= \braket{i (d_{k\sigma}^\dagger h_{\sigma\tau'}(k)) d_{k\tau} + eE\, d_{k\tau'}^\dagger \frac{\partial}{\partial k} d_{k\tau} + h.c.} \\
    &= i \left( \braket{d_{k\sigma}^\dagger d_{k\tau}} h_{\sigma\tau'}(k) - h.c. \right) + eE \braket{\frac{\partial}{\partial k} (d_{k\tau'}^\dagger d_{k\tau})} \\
    &= -i \left[\rho(k, t), h(k)\right]_{\tau\tau'} + eE \frac{\partial \rho_{\tau\tau'}}{\partial k}.
\end{align}
Rearranging and replacing $\rho \to f$ gives the Boltzmann equation \eqref{eq:Boltzmann} without the collision integral.

To transform the spin-space \eqref{eq:Boltzmann} into its band-space equivalent \eqref{eq:Boltzmann_bandspace}, we multiply by $\mathcal U^\dagger(k)$ and $\mathcal U(k)$ from the left and right, respectively, insert the definitions of $\tilde f$ and $\tilde I[\tilde f]$, and use that
\begin{gather}
    \frac{\partial}{\partial k} (\mathcal U \tilde f \mathcal U^\dagger) = \frac{\partial \mathcal U}{\partial k} \tilde f \mathcal U^\dagger + \mathcal U \tilde f\, \frac{\partial\mathcal U^\dagger}{\partial k} + \mathcal U \frac{\partial \tilde f}{\partial k}\, \mathcal U^\dagger,
\end{gather}
so
\begin{align}
    \mathcal U^\dagger \frac{\partial}{\partial k} (\mathcal U \tilde f \mathcal U^\dagger)\, \mathcal U &= -i \mathcal A \tilde f + i\tilde f \mathcal A + \frac{\partial\tilde f}{\partial k} \\
    &= -i [\mathcal A, \tilde f] + \frac{\partial\tilde f}{\partial k}
\end{align}
with the non-abelian Berry connection
\begin{align}
    \mathcal A &= i\mathcal U^\dagger \frac{\partial\mathcal U}{\partial k} = -i \frac{\partial \mathcal U^\dagger}{\partial k}\, \mathcal U \\
    &= -\frac{J_{sd}^2 u a}{\sqrt 2 \left( J_{sd^2} + 2u^2 \sin^2(k a) \right)^{3/2}}\, \sigma_y \\
\label{eq:SM_Berry_conn_approx}
    &\overset{u \ll J_{sd}}{\longrightarrow} - \frac{u a}{\sqrt 2\, J_{sd}}\, \sigma_y.
\end{align}
Since $\mathcal A \equiv \mathcal A_y \sigma_y$ is proportional to $\sigma_y$ and thus off-diagonal in the band indices, it is responsible for interband tunneling.

Note that, in the absence of band crossings, $\mathcal A$ is a $U(1) \times U(1)$ gauge connection since the diagonalization matrix $\mathcal U \in U(2)$ has two $U(1)$ gauge degrees of freedom corresponding to a $k$-dependent change of the relative and absolute phases of the two eigenvectors, respectively.
A $U(1) \times U(1)$ gauge transformation $G(k) = e^{i(\chi_k \mathbbm 1 + \varphi_k \sigma_z)}$ acts as $\mathcal U \to \mathcal U G$ and $\mathcal A \to G^\dagger \mathcal A G + i G^\dagger \partial G/\partial k$.

\sectionPRL{\thesection.~Magnetization and charge current}
\label{sec:SM_magn_and_charge_curr}
We obtain the magnetization phase-space density $\vec m(k, t)$ from the band-space distribution function $\tilde f$ by first transforming the physical spin operator $\vec\sigma/2$ into the rotated spin basis of the $d$-fermions using the matrices $R_j$ in \eqref{eq:SM_spin_rot_mat}; then transforming it into the band basis using the matrix $\mathcal U$; and finally averaging over the four-site unit cell:
\begin{align}
    \vec m(k, t) &= \frac{1}{4} \sum_{j=1}^4 \operatorname{Tr} \left( \tilde f(k, t)\, \mathcal U^\dagger(k) R_j \frac{\vec\sigma}{2} \mathcal U(k) \right).
\end{align}
Using that $(1/4) \sum_{j=1}^4 R_j \vec\sigma = (0, 0, \sigma_y)^T$, we find that the $x$- and $y$-components of $\vec m$ vanish.
The $z$-component is
\begin{align}
    m_z(k, t) &= \frac{1}{2} \operatorname{Tr} \left( \tilde f(k, t)\, \mathcal U^\dagger(k) \sigma_y\, \mathcal U(k) \right) \\
    &= \frac{1}{2} \operatorname{Tr} \left( \tilde f(k, t)\, (\cos\vartheta_k \sigma_x + \sin\vartheta_k \sigma_z) \right).
\end{align}
Decomposing $\tilde f = \tilde f_0 \mathbbm 1 + \vec{\tilde f} \cdot \vec\sigma$ into a scalar unpolarized component $f_0$ and a vectorial spin-polarized component $\vec f$, we obtain
\begin{align}
\label{eq:SM_magn_dens}
    m_z(k, t) = \tilde f_x \cos\vartheta_k + \tilde f_z \sin\vartheta_k.
\end{align}
The total magnetization in $z$-direction is $M_z(t) = \int_{-\pi/a}^{\pi/a} a \mathrm dk/(2\pi)\, m_z(k, t)$.

The charge current phase-space density $j(k, t)$ is obtained in terms of the group velocity matrix $\partial h(k)/\partial k$ as
\begin{align}
    &j(k, t) = -e \operatorname{Tr} \left( f(k, t)\, \frac{\partial h(k)}{\partial k} \right) \\
    &= -e \operatorname{Tr} \left( \mathcal U(k) \tilde f(k, t)\, \mathcal U^\dagger(k) \frac{\partial (\mathcal U(k) D(k)\, \mathcal U^\dagger(k))}{\partial k} \right) \\
    &= -e \operatorname{Tr} \left( \tilde f(k, t)  \frac{\partial D(k)}{\partial k} \right) \\
    &\quad + i e \operatorname{Tr} \left( \tilde f(k, t) \left[ \mathcal A(k), D(k) \right] \right) \\
    &\equiv j_\mathrm{band}(k, t) + j_\mathrm{tunnel}(k, t).
\end{align}
The current has two contributions: the pure intraband transport current $j_\mathrm{band}(k, t)$, which only depends on the group velocities $\partial\epsilon_\pm(k)/\partial k$ of the individual bands, and the interband tunneling current $j_\mathrm{tunnel}(k, t)$ which depends on the non-abelian Berry connection.
Note that $j_\mathrm{tunnel}$ is invariant under a $U(1) \times U(1)$ gauge transformation $G(k)$ of the eigenvector phases because $G$, $\partial G/\partial k$, and $D(k)$ are diagonal matrices (cf.\ the transformation behavior of $\mathcal A$ discussed in Sec.~\ref{sec:SM_Boltzmann_eq} of the SM).
Splitting up the band dispersion $\epsilon_\pm(k) = \epsilon_0(k) \pm \epsilon(k)$ with $\epsilon_0(k) = -\sqrt 2\, u \cos(k a)$ and $\epsilon(k) = (J_{sd}^2 + 2u^2\sin^2(k a))^{1/2}$, we find that
\begin{align}
    j_\mathrm{band}(k, t) &= -2e \left( \tilde f_0(k, t) \frac{\partial \epsilon_0}{\partial k} + \tilde f_z(k, t) \frac{\partial \epsilon}{\partial k} \right), \\
    j_\mathrm{tunnel}(k, t) &= -2e \tilde f_x(k, t) \mathcal A_y(k).
\end{align}
The total current is $J(t) = \int_{-\pi/a}^{\pi/a} a\,\mathrm dk/(2\pi)\, j(k, t) \equiv J_\mathrm{band}(t) + J_\mathrm{tunnel}(t)$.
After integration by parts, we get
\begin{align}
\label{eq:SM_curr}
    J_\mathrm{band}(t) &= 2e \int_{-\pi/a}^{\pi/a} \frac{a\, \mathrm dk}{2\pi} \left( \epsilon_0(k) \frac{\partial\tilde f_0}{\partial k} + \epsilon(k) \frac{\partial\tilde  f_z}{\partial k} \right).
\end{align}

\sectionPRL{\thesection.~Analytic solutions with static electric field}
\label{sec:SM_static_field}
If the band gap is large compared to the band width and the Bloch frequency, i.e., $J_{sd} \gg u, \omega_B$, we may neglect the $\mathcal A$-dependent intraband tunneling term in \eqref{eq:Boltzmann_bandspace}.
In a sense, the acceleration by the electric field ``slow enough" to prevent intraband transitions, which leads us to call this the \textit{adiabatic} approximation.
In fact, we later show that the leading interband tunneling correction of the distribution function is of order $u\omega_B/J_{sd}^2$ (see \eqref{eq:SM_nonadiabatic_corr}), so either condition $u \ll J_{sd}$ or $\omega_B/J_{sd}$ would be sufficient to suppress interband tunneling.
However, we use the stronger condition  $u,\omega \ll J_{sd}$ for our ``adiabatic limit" as it is analytically more convenient and suppresses interband transitions even more strongly.
We later consider the impact of higher-order corrections in $u/J_{sd}$ at zeroth order in $\omega_B/J_{sd}$ (see \eqref{eq:SM_magn_HH} and \eqref{eq:SM_curr_HH}). 

Assuming that the chemical potential lies fully inside the lower band, the initial equilibrium condition is $\tilde f(k, t=0) = \tilde f^\mathrm{eq}(k) = \Theta(\mu - \epsilon_-(k)) (\mathbbm 1 - \sigma_z)/2$.
This leads to two decoupled Boltzmann equations for the diagonal components of $\tilde f$, while the off-diagonal components of $\tilde f$ (i.e., $\tilde f_x$ and $\tilde f_y$) remain zero at all times.
Now, also the term $i[D(k), \tilde f]$ vanishes because both $\tilde f$ and $D(k)$ are diagonal and commute.
In matrix form, the approximated Boltzmann equation is
\begin{align}
    \frac{\partial\tilde f}{\partial t} - eE\frac{\partial\tilde f}{\partial k} = -\frac{1}{\tau} (\tilde f - \tilde f^\mathrm{eq}).
\end{align}
This inhomogeneous partial differential equation can be solved, e.g., with the method of Green's functions.
The relevant Green's function in this case is 
\begin{align*}
    \mathcal G(k,t; k',t') = e^{-(t-t')/\tau} \Theta(t-t') \, \delta\!\left( k - k' + eE(t - t') \right),
\end{align*}
leading to the distribution function
\begin{align}
    \tilde f(k, t) &= e^{-t/\tau} \tilde f^\mathrm{eq}(k + eEt) \\
\label{eq:SM_distfct_inhom_sol}
    &\quad + \frac{1}{\tau} \int_0^t \mathrm dt^\prime\, e^{-t^\prime/\tau} \tilde f^\mathrm{eq}(k + eEt^\prime).
\end{align}
To compute the magnetization to leading order in $u/J_{sd}$, we use that $\tilde f_x$ in \eqref{eq:SM_magn_dens} vanishes at all times as established above, and we expand $\sin\vartheta_k \approx \sqrt 2\, (u/J_{sd}) \sin(ka)$.
It is convenient to first evaluate the momentum integral over the Brillouin zone, and only then the time integral of \eqref{eq:SM_distfct_inhom_sol}.
For the charge current, we use \eqref{eq:SM_curr} alongside the fact that $\epsilon_-(k) \approx -\sqrt 2\,u \cos(ka) - J_{sd}$ to leading order in $u/J_{sd}$.
The result is
\begin{align}
\label{eq:SM_magn_adiabatic}
    M_z(t) &= \frac{1}{\sqrt 2\,\pi} \frac{u}{J_{sd}} \sin(k_F a)\, \Gamma(t, \omega_B), \\
    \label{eq:SM_curr_adiabatic}
    J(t) &= \frac{\sqrt 2}{\pi}\, eu \sin(k_Fa)\, \Gamma(t, \omega_B)
\end{align}
where $k_F \ge 0$ is the Fermi momentum, defined by $\epsilon_-(k_F) = \mu$, and
\begin{widetext}
\begin{align}
\label{eq:SM_Gamma}
    \Gamma(t, \omega_B) \equiv \frac{\omega_B\tau}{1 + (\omega_B\tau)^2} + e^{-t/\tau} \left( \sin(\omega_B t) + \frac{1}{1 + (\omega_B\tau)^2} \left( \omega_B \tau \cos(\omega_B t) + \sin(\omega_B t) \right) \right).
\end{align}
\end{widetext}
\eqref{eq:SM_magn_adiabatic} and \eqref{eq:SM_curr_adiabatic} are used to plot the dashed lines in Fig.~\ref{fig:fig1}(d-g).
In the limit of a pristine crystal without scattering, $\omega_B\tau \gg 1$, we obtain the results given in the main text.
If the scattering time $\tau$ is finite, taking the  long-time limit $t \gg \tau$ leads to the finite Edelstein magnetization stated in the main text.

\begin{figure*}
    \centering
    \includegraphics[width=0.9\linewidth]{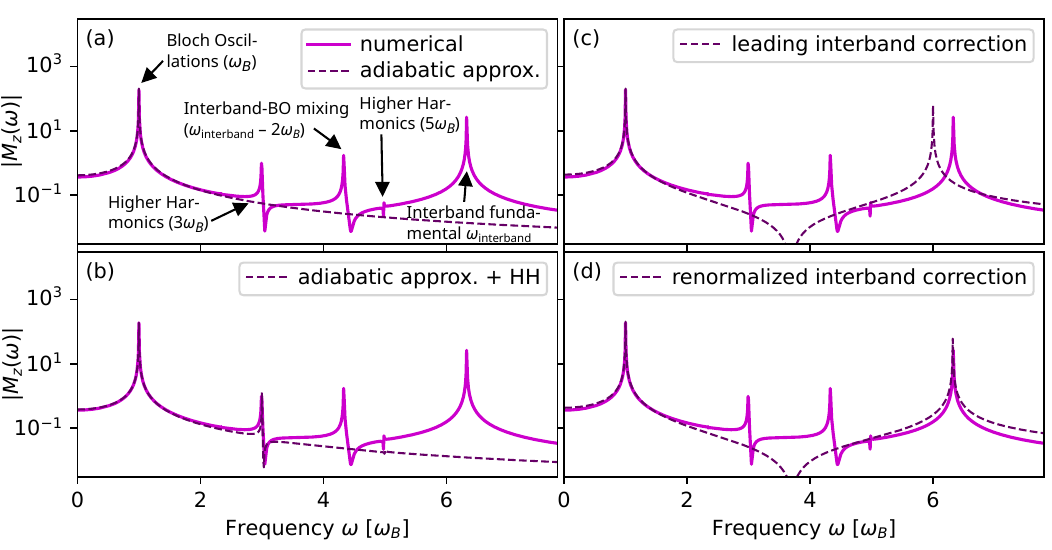}
    \caption{Fourier spectra of the magnetization for $\mu = -3u$, $J_{sd} = 3u$, $u=\omega_B$, and $\tau = \infty$.
    The numerical solution (solid lines) is the same in all panels.
    To regularize the Fourier transform in the $\tau=\infty$ limit, the real-time magnetization was multiplied by an exponential decay factor with characteristic decay time $1000/u$ (note that this is different from a finite relaxation time $\tau=1000/u$ because the latter would produce the Edelstein effect).
    The analytical solutions (dashed lines) are obtained within different approximations:
    (a) the adiabatic approximation, which only takes into account $u/J_{sd}$ and $\omega_B/J_{sd}$ at lowest order, and therefore only reproduces the fundamental Bloch oscillation (BO) frequency;
    (b) the adiabatic approximation with higher harmonics of the Bloch frequency $\omega_B$, which takes into account all orders in $u/J_{sd}$ but only the lowest order in $\omega_B/J_{sd}$, and therefore does not capture the interband tunneling;
    (c) including interband tunneling corrections to the adiabatic approximation at lowest order in $u/J_{sd}$ but next-to-lowest order in $\omega_B/J_{sd}$; and
    (d) the same approximation as in panel (c) except with the renormalized interband tunneling frequency, which is exact to all orders in $u/J_{sd}$, but not including any other effects of higher orders in $u/J_{sd}$ such as interband-BO mixing.}
    \label{fig:figSM1}
\end{figure*}

The higher harmonics of $\omega_B$ observed in the numerical Fourier spectra in Fig.~\ref{fig:fig1}(f-g) can be analytically understood in the limit $\tau\to\infty$ by keeping the higher order contributions in $u/J_{sd}$, which describe the deviations of the band dispersion from a pure cosine, while still enforcing the limit $\omega_B \ll J_{sd}$ to suppress interband tunneling.
Concretely, this amounts to keeping the full expressions for $\sin\vartheta_k$ and $\epsilon_-(k)$ in the computation of the magnetization and the current, respectively.
We find
\begin{widetext}
\begin{align}
\label{eq:SM_magn_HH}
    M_z(t) &= \frac{1}{4\pi} \left( \arcsin\left( C \cos(k_F a - \omega_B t) \right)  - \arcsin\left( C \cos(k_F a + \omega_B t) \right)\right), \\
\label{eq:SM_curr_HH}
    J(t) &= \frac{e}{2\pi} \left( 2\sqrt 2\,u \sin(k_F a) \sin(\omega_B t) - \sqrt{J_{sd}^2 + 2u^2 \sin^2(k_F a - \omega_B t)} + \sqrt{J_{sd}^2 + 2u^2 \sin^2(k_F a + \omega_B t)} \right).
\end{align}
\end{widetext}
where $C \equiv \sqrt{2u^2/(J_{sd}^2 + 2u^2)}$.
Expanding these expressions in $u/J_{sd}$ and applying the relevant trigonometric identities reveals that, in this minimal model, the magnetization (charge current) only contains odd (even) higher harmonics of $\omega_B$.
The Fourier spectrum of the magnetization in \eqref{eq:SM_magn_HH} is depicted in Fig.~\ref{fig:figSM1}(b) and shows good agreement with the numerical result.
Note that this higher harmonic spectrum is not universal.
For example, including a simple next-nearest neighbor hopping can give rise to odd higher harmonics in the current.

Interestingly, in the minimal model the magnetization exhibits no components oscillating at frequency $4\omega_B = eE \cdot (4a)$, in contrast to what one might expect from the four-site unit cell.
This is \textit{not} an artifact of the transformation from $c$ to $d$-fermions, which reduces the four-site unit cell down to one, or any of our approximations.
Instead, it is enforced by the $[C_4^z || \vec t_1]$ symmetry of the model, which forbids gap openings at the momenta $\pi/(4a)$, $\pi/(2a)$, and $3\pi/(4a)$.

Finally, we consider the leading order interband tunneling correction to the adiabatic approximation, which enters at linear order in $\omega_B/J_{sd}$.
For simplicity, we restrict the analysis to infinite relaxation time limit $\tau \to \infty$.
With interband tunneling, the off-diagonal components of the distribution function $\tilde f$ are not zero anymore.
Therefore, the term $i[D(k),\tilde f]$ in the Boltzmann equation \eqref{eq:Boltzmann_bandspace} generally does not vanish.
Since the non-abelian Berry connection $\mathcal A$ is of order $u/J_{sd}$, we treat the term $ieE[\mathcal A, \tilde f]$ in the Boltzmann equation as a perturbation.
Then, the Boltzmann equation takes the form
\begin{align}
\label{eq:SM_Boltzmann_nonadiabatic}
    \frac{\partial\tilde f}{\partial t} - eE \frac{\partial\tilde f}{\partial k} + i [D(k), \tilde f] = -ieE [\mathcal A(k), \tilde f],
\end{align}
which can be formally solved via the method of Green's functions by treating the right hand side as an inhomogeneity.
The relevant Green's function of the (matrix-valued) operator $\left(\partial/\partial t - eE\partial/\partial k + i[D(k), \cdot]\right)_{nm}$ with band indices $n,m$ is given by
\begin{align*}
    &\mathcal G(k,t,nm; k',t',n'm') \\
    &= \Theta(t - t')\, \delta\! \left( k - k' + eE(t - t') \right) \! V_{nn'}(k,k') V_{mm'}^*(k,k'),
\end{align*}
where $V_{nn'}(k,k') = \exp((i/eE) \! \int_{k'}^k \mathrm dp\, \epsilon_n(p)) \delta_{nn'}$ and $\epsilon_n(p) = D_{nn}(p)$ denotes the dispersion of the upper ($n=+$) and lower ($n=-$) band.
In total, this leads to the following self-consistent expression for the interband tunneling corrections $\delta\tilde f$ to the distribution function:
\begin{widetext}
\begin{align}
    \delta\tilde f_{nm}(k,t) = -i \int_k^{k+eEt} \mathrm dk'\, e^{\frac{i}{eE} \! \int_{k'}^k \mathrm dp\, (\epsilon_n(p) - \epsilon_m(p))} \left[\mathcal A(k'), \tilde f\left(k', t + \frac{k - k'}{eE}\right)\right]_{nm}.
\end{align}
\end{widetext}
To first order in $\omega_B/J_{sd}$, we replace the exact distribution function with its adiabatic solution, 
$\tilde{f}(k, t) \approx \tilde{f}^\mathrm{eq}(k + eEt) = \Theta(\mu - \epsilon_-(k+eEt)) (1 - \sigma_z)/2$.
To simplify the calculation further, we additionally assume the limit $u \ll J_{sd}$, in which case the Berry connection is given by \eqref{eq:SM_Berry_conn_approx}.
Decomposing $\delta\tilde f = \delta\tilde f_0 \mathbbm 1 + \delta\vec{\tilde f} \cdot \vec\sigma$, we find that $\delta\tilde f_{0} = 0$, and that
\begin{widetext}
\begin{align}
\label{eq:SM_nonadiabatic_corr}
    \delta\vec{\tilde f}(k, t) = \frac{u\omega_B}{\sqrt 2\,J_{sd}^2} \Theta(\mu - \epsilon_-(k+eEt)) \int_0^{eEt} \mathrm dk' \begin{pmatrix}
        \cos\left( \frac{2}{eE} \int_0^{k'-k} \mathrm dp\, \epsilon(p) \right) \\
        \sin\left( \frac{2}{eE} \int_0^{k'-k} \mathrm dp\, \epsilon(p) \right) \\
        0
    \end{pmatrix},
\end{align}
\end{widetext}
where $\epsilon(p) = (J_{sd}^2 + 2u^2\sin^2(ka))^{1/2}$.
The remaining integrals are solvable in the limit $u \ll J_{sd}$, where the bands are approximately equidistant, $\epsilon(p) = J_{sd} + \mathcal O(u^2/J_{sd})$.
Then, the correction to the magnetization at leading order in $\omega_B/J_{sd}$ and $u/J_{sd}$ according to \eqref{eq:SM_magn_dens} and \eqref{eq:SM_nonadiabatic_corr} is 
\begin{align}
    \delta M_z(t) = \frac{u\omega_B k_F}{2\sqrt 2\,\pi J_{sd}^2} \sin(2J_{sd} t),
\end{align}
where we approximated $\cos\vartheta_k \approx 1 + \mathcal O(u/J_{sd})$.
We observe that interband tunneling leads to additional oscillations at the fundamental frequency $\omega_\mathrm{interband} \equiv 2J_{sd} + \mathcal O(u^2/J_{sd})$, which roughly agrees with the numerical Fourier spectra even though a small deviation is visible (see Fig.~\ref{fig:figSM1}(c)).
Physically, a small fraction of the electron population continuously tunnels between the lower and the upper bands at a frequency that corresponds (approximately) to the band gap, similar to Larmor precession in a two-state quantum system.
Since the lower and the upper band have the opposite spin polarization, the tunneling electrons flip their spins twice during each oscillation period, manifesting as oscillations at frequency $\omega_\mathrm{interband}$ in the magnetization.
The deviation of $2J_{sd}$ from the numerical interband tunneling frequency is explained by higher order corrections in $u/J_{sd}$, which renormalize $\omega_\mathrm{interband}$ and give rise to sum and difference frequency oscillations of $\omega_\mathrm{interband}$ and $\omega_B$ (interband-BO mixing).
It is possible to obtain the renormalized fundamental frequency non-perturbatively to all orders in $u/J_{sd}$.
First, we note that the inner integrals in \eqref{eq:SM_nonadiabatic_corr} have the form
\begin{align}
    \int_0^{k'-k} \mathrm dp\, \epsilon(p) &= C (k'-k) + \mathrm{osc.},
\end{align}
where $C$ is some constant and $\mathrm{osc.}$ denotes oscillatory terms in $k'-k$ with period $\pi/a$.
The latter contribute to the sum and difference frequency signals, while the linear term in $k'-k$ yields the renormalization of $\omega_\mathrm{interband}$.
Disregarding the oscillatory terms and noting that $C = \int_{0}^{\pi/a} (a\,\mathrm dp/\pi)\, \epsilon(p)$ is just the average of the integrand over one period, we obtain
\begin{align}
    \omega_\mathrm{interband} = 2 C &= \frac{4}{\pi} J_{sd} E\!\left( -\frac{2u^2}{J_{sd}^2} \right) \\
    &\approx 2J_{sd} + \frac{u^2}{J_{sd}} + \dots
\end{align}
with $E(m) = \int_0^{\pi/2} \mathrm ds\, (1 - m \sin^2(s))^{1/2}$ the complete elliptic integral of the second kind.
This result is in perfect agreement with the numerical frequency of the interband tunneling peak, see Fig.~\ref{fig:figSM1}(d).
We expect higher orders in $\omega_B/J_{sd}$ to yield higher harmonics of the interband tunneling frequency.

\sectionPRL{\thesection.~Analytic solutions with oscillating electric field}
\label{sec:SM_oscillating_field}
With an oscillating field $E(t) = E_0 \cos(\Omega t)$, the coefficient in front of the momentum derivative in the Boltzmann equation becomes time-dependent.
In the adiabatic approximation, the homogeneous part of the Boltzmann equation is solved by
\begin{align*}
    \tilde f^\mathrm{eq}\left(k + e \int_0^t \mathrm dt' E(t')\right)
    = \tilde f^\mathrm{eq}\left(k + \frac{eE_0}{\Omega} \sin(\Omega t) \right)
\end{align*}
To find a particular solution, we again employ the method of Green's functions.
The relevant Green's function is given by
\begin{align*}
    &\mathcal G(k,t; k',t') \\
    &= e^{-(t - t')/\tau} \Theta(t-t')\, \delta\!\left( k - k' + e \int_{t'}^t \mathrm dt'' E(t'') \right) 
%
\end{align*}
where $\int_{t'}^t \mathrm dt'' E(t'') = (E_0/\Omega) (\sin(\Omega t) - \sin(\Omega t'))$.
Note that, since the coefficients of the Boltzmann equation are now time-dependent, the Green's function is no longer invariant under time translations.
The full inhomogeneous solution of the Boltzmann equation reads
\begin{align}
    &\tilde f(k, t) = \tilde f^\mathrm{eq}\left(k + \frac{eE_0}{\Omega} \sin(\Omega t) \right) + \frac{1}{\tau} \int_0^t \mathrm dt'  \\
    & \quad \cdot e^{-(t - t')/\tau} \tilde f^\mathrm{eq} \left( k + \frac{eE_0}{\Omega} (\sin(\Omega t) - \sin(\Omega t')) \right)
\end{align}
The magnetization and the charge current in the adiabatic approximation (i.e., to leading order in $u/J_{sd}$) are given by
\begin{align}
\label{eq:SM_magn_oscil_field}
    M_z(t) &= \frac{1}{\sqrt 2\,\pi} \frac{u}{J_{sd}} \sin(k_F a)\, \Delta(t), \\
    \label{eq:SM_curr_oscil_field}
    J(t) &= \frac{\sqrt 2}{\pi}\, eu \sin(k_F a)\, \Delta(t),
\end{align}
where
\begin{widetext}
\begin{align}
\label{eq:SM_Delta}
    \Delta(t) = \sin\left( \frac{eE_0 a}{\Omega} \sin(\Omega t) \right) + \frac{1}{\tau} \int_0^t \mathrm dt'\, e^{-(t - t')/\tau} \sin\left( \frac{eE_0 a}{\Omega} (\sin(\Omega t) - \sin(\Omega t')) \right).
\end{align}
\end{widetext}
The remaining time integral cannot be evaluated analytically.
We therefore compute it numerically using the SciPy function \texttt{integrate.quad}, yielding the dashed lines shown in Fig.~\ref{fig:fig2}(d-f).
We note that our formalism is equivalent to the one used in Ref.~\cite{vampa_theoretical_2014}.

To elucidate the meaning of our result, we consider some limits.
First, when $\tau \to \infty$, the time integral in \eqref{eq:SM_Delta} does not contribute and we are only left with the forever oscillating term $\sin(eE_0a/\Omega \sin(\Omega t))$.
\begin{enumerate}
    \item If the electric field amplitude is furthermore weak, $eE_0 a \ll \Omega$, the outer sine function can be expanded, yielding magnetization and current that oscillate only at the driving frequency $\Omega$, as expected within linear response theory.
    Physically, the field induces only a small deviation of the electrons from their equilibrium distribution, preventing them from crossing the Brillouin zone boundary and thus suppressing any signatures of Bloch oscillations.
    
    \item If, on the contrary, the oscillation frequency is small, $\Omega \ll eE_0 a$, the inner sine function can instead be expanded, yielding oscillations at the Bloch frequency $eE_0a$.
    Physically, the electric field is then nearly static and can accelerate electrons across the Brillouin zone boundary, giving rise to Bloch oscillations in the same way as a static electric field.
    
    \item For intermediate field strengths, $eE_0 a \sim \Omega$, we can access the frequency spectrum of $M_z$ and $J$ via the Jacobi-Anger expansion $\sin(z \sin(\Omega t)) = 2 \sum_{n\in\mathbb N\text{ odd}} J_n(z) \sin(n\Omega t)$, where $J_n$ is the $n$-th Bessel function of the first kind.
    The acceleration of the electrons beyond the Brillouin zone boundaries generates odd higher harmonics of the driving frequency, as observed in \LightwaveElectronics{} spectroscopy.
    This phenomenon goes by the name of higher-harmonic generation (HHG), as discussed in the main text.
    The amplitudes of the corresponding HHG frequency peaks are weighted by the factor $J_n(z)$, which, for fields $eE_0a \lesssim \sqrt{n}\,\Omega$, is well approximated by $z^n \propto E_0^n$.
    Instead of the amplitudes, experiments typically measure the intensities of the HHG peaks.
    These are given by the squared amplitudes, yielding the established perturbative scaling law $\propto E_0^{2n}$ for the intensities~\cite{schmid_tunable_2021}.
    
\end{enumerate}
When $\tau$ is finite, the time integral cannot be ignored.
Although a closed-form solution is not available, we can analyze its effect on the frequency spectrum of $M_z$ and $J$.
To this end, we write the outer sine function as $\sin(\dots) = \operatorname{Im} e^{i \dots}$ and use the Jacobi-Anger expansion $e^{i z \sin(\Omega t)} = \sum_{n\in\mathbb Z} J_n(z)\, e^{in\Omega t}$ with $z = eE_0 a/\Omega$:
\begin{align*}
    \int_0^t \mathrm dt' \dots
    &= \operatorname{Im} \sum_{n,m\in\mathbb Z} \! J_n J_m \, e^{(-\frac{1}{\tau} + im\Omega) t} \int_0^t \frac{\mathrm dt'}{\tau}\, e^{(\frac{1}{\tau} - in\Omega) t'} \\
    &= \operatorname{Im} \sum_{n,m\in\mathbb Z} \! J_n J_m \, e^{(-\frac{1}{\tau} + im\Omega) t} \, \frac{e^{(\frac{1}{\tau} - in\Omega) t} - 1}{1 - in\Omega\tau} \\
    &= \operatorname{Im} \sum_{n,m\in\mathbb Z} \! J_n J_m  \frac{e^{i(m-n)\Omega t} - e^{-t/\tau} e^{im\Omega t}}{1 - in\Omega\tau},
\end{align*}
where we suppressed the $z$-dependence of the Bessel functions for brevity.
Introducing the phase $\varphi_n \equiv \operatorname{arg}(1 + n\Omega t) = -\varphi_{-n}$, we find that
\begin{align*}
    &\int_0^t \mathrm dt' \dots \\
    &= \operatorname{Im} \sum_{n,m\in\mathbb Z} \! J_n J_m \frac{e^{i((m-n)\Omega t + \varphi_n)} - e^{-t/\tau} e^{i(m\Omega t + \varphi_n)}}{\sqrt{1 + (n\Omega\tau)^2}} \\
    &= \sum_{n,m\in\mathbb Z} \! J_n J_m \frac{\sin((m-n)\Omega t + \varphi_n) - e^{-t/\tau} \sin(m\Omega t + \varphi_n)}{\sqrt{1 + (n\Omega\tau)^2}},
\end{align*}
Due to the relation $\varphi_n = -\varphi_{-n}$ and the property $J_{-n} = (-1)^n J_n = (-1)^{-n} J_n$ of the Bessel functions, only the summands with odd $m-n$ are nonzero.
This implies that the term $\sin((m-n)\Omega t + \varphi_n)$ only ever generates odd higher harmonics of the driving frequency.
However, the term $\sin(m\Omega t + \varphi_n)$ can also generate \textit{even} higher harmonics, as shown in Fig.~\ref{fig:fig2}(a).
Due to the prefactor $e^{-t/\tau}$, these harmonics are broadened, with a linewidth of $1/\tau$.
Furthermore, their amplitudes are suppressed by the factor of $1/\tau$ in front of the time integral in \eqref{eq:SM_Delta}, which explains the small magnitude of the even HHG peaks in Fig.~\ref{fig:fig2}(a).

\sectionPRL{\thesection.~Analytic solutions for the 2D model}
\label{sec:SM_2D_model}
In the 2D case, the Bloch Hamiltonian $h(\vec k)$, the dispersion $\epsilon_\pm(\vec k)$, and the quantity $\sin\vartheta_{\vec k}$ are identical to the 1D model, except that all occurrences of $\sin(ka)$ and $\cos(ka)$ are replaced with $S(\vec k) \equiv \sin(k_x a) + \sin(k_y a)$ and $C(\vec k) \equiv \cos(k_x a) + \cos(k_y a)$, respectively.
In the Boltzmann equation, the momentum derivative term gets modified into $e\vec E \cdot \partial f(\vec k,t)/\partial \vec k$.
The expressions for the magnetization and the charge current are identical to the 1D model, except that the momentum integration is performed over the 2D Brillouin zone of the square lattice.
Since the calculations are more involved, we focus only on the limit of infinite relaxation time $\tau\to\infty$ in the adiabatic approximation.
Employing \eqref{eq:SM_magn_dens}, approximating $\sin\vartheta_{\vec k} \approx \sqrt 2\, (u/J_{sd})\, S(\vec k)$ to leading order in $u/J_{sd}$, and using the solution of the Boltzmann equation $\tilde f(\vec k,t) = \tilde f^\mathrm{eq}(\vec k + e\vec E t)$ with the equilibrium distribution $\tilde f^\mathrm{eq}(\vec k, t) = \Theta(\mu - \epsilon_-(\vec k)) (\mathbbm 1 - \sigma_z)/2$, we obtain
\begin{align}
    &M_z(t) = \int_\mathrm{BZ}  f_z(\vec k, t) \sin\vartheta_{\vec k} \\
    &\approx -\frac{u}{\sqrt 2\, J_{sd}} \int_\mathrm{BZ} \Theta(\mu - \epsilon_-(\vec k + e\vec E t))\, S(\vec k),
\end{align}
where $\int_\mathrm{BZ} \equiv \int_{-\pi/a}^{\pi/a} \! \frac{a\, \mathrm dk_x}{2\pi} \! \int_{-\pi/a}^{\pi/a} \! \frac{a \mathrm dk_y}{2\pi}$.
Shifting $\vec k \to \vec k - e\vec E t$ in the integrals, approximating $\epsilon_-(\vec k) \approx -\sqrt 2\, u\, C(\vec k) - J_{sd}$ to leading order in $u/J_{sd}$, and introducing the shorthand $\lambda \equiv (\mu + J_{sd}) / (\sqrt 2\,u)$, we find
\begin{align}
    M_z(t) = -\frac{u}{\sqrt 2\, J_{sd}} \int_\mathrm{BZ} \Theta(\lambda - C(\vec k))\, S(\vec k - e \vec E t).
\end{align}
Next, we apply the trigonometric addition formula to the sines and use that odd terms in $\vec k$ integrate to zero, leading to
\begin{align}
    M_z(t) &= \frac{u \sin(eE_x t)}{\sqrt 2\, J_{sd}}  \int_\mathrm{BZ} \Theta(\lambda - C(\vec k)) \cos(k_x a) \\
    &\quad + (x \leftrightarrow y).
\end{align}
To solve the two-dimensional integral, we substitute $c_{x/y} = \cos(k_{x/y} a)$, integrate by parts over $c_x$, thereby turning the $\Theta$-function into a $\delta$-function, evaluate the $c_x$ integral using that $\delta$-function, and massage the remaining $c_y$ integral into an expression depending only on the complete elliptic integrals $K(m) = \int_0^1 \mathrm ds/\left( (1 - s^2) (1 - ms^2) \right)^{1/2}$ and $E(m) = \int_0^1 \mathrm ds \left( (1 - ms^2) / (1 - s^2) \right)^{1/2}$, where $m \equiv [(2 - |\lambda|)/(2 + |\lambda|)]^2$.
The calculation of the charge current $\vec J(t)$ proceeds analogously.
The final result is
\begin{align}
\label{eq:SM_magn_adiabatic_2D}
    M_z(t) &= \frac{\sqrt 2\, u}{\pi^2 J_{sd}} (\sin(\omega_{B,x} t) + \sin(\omega_{B,y} t))\, \Lambda(\lambda), \\
\label{eq:SM_curr_adiabatic_2D}
     \vec J(t) &= \frac{2\sqrt 2\, eu}{\pi^2} \begin{pmatrix}
         \sin(\omega_{B,x} t) \\ \sin(\omega_{B,x} t)
     \end{pmatrix} \Lambda(\lambda),
\end{align}
where $\omega_{B,x/y} = eE_{x/y} a$ are the Bloch frequencies in $x$- and $y$-direction, and
\begin{widetext}
\begin{align}
    \Lambda(\lambda) = \begin{cases}
        -|\lambda| K(m) + \left( 1 + \frac{|\lambda|}{2} \right) E(m) & -2 < \lambda < 2 \\
        0 & \text{otherwise.}
    \end{cases}
\end{align}
\end{widetext}
To include a finite relaxation time, the $\sin(\omega_{B,x/y} t)$ terms in \eqref{eq:SM_magn_adiabatic_2D} and \eqref{eq:SM_curr_adiabatic_2D} need to be replaced with $\Gamma(t, \omega_{B,x/y})$ as defined in \eqref{eq:SM_Gamma}.

\end{document}